\documentclass[journal=jpccck,manuscript=article]{achemso}

\usepackage[version=3]{mhchem} % Formula subscripts using \ce{}

\usepackage{color}
\usepackage{graphicx}
\usepackage{caption}
\usepackage{subcaption}
\usepackage{comment}

\author{Juan Pablo Guerrero-Felipe}
\affiliation{Physics Department and IRIS Adlershof, Humboldt-Universität zu Berlin, 12489 Berlin, Germany}
\alsoaffiliation{Department of Physics, Freie Universit{\"a}t Berlin, 14195 Berlin, Germany}
\author{Ana M. Valencia}
\affiliation{Institute of Physics, Carl-von-Ossietzy Universit{\"a}t Oldenburg, 26129 Oldenburg, Germany}
\alsoaffiliation{Physics Department and IRIS Adlershof, Humboldt-Universität zu Berlin, 12489 Berlin, Germany}
\author{Caterina Cocchi}
\affiliation{Physics Department and IRIS Adlershof, Humboldt-Universität zu Berlin, 12489 Berlin, Germany}
\alsoaffiliation{Institute of Physics, Carl-von-Ossietzy Universit{\"a}t Oldenburg, 26129 Oldenburg, Germany}
\alsoaffiliation{Center for Nanoscale Dynamics (CeNaD), Carl-von-Ossietzy Universit{\"a}t Oldenburg, 26129 Oldenburg, Germany}
\email{caterina.cocchi@uni-oldenburg.de}

\title{Magnification of Plasmon Resonances in Monolayer MoS$_{2}$ via Conjugated Molecular Adsorbates}

\begin{document}

\begin{tocentry}

\includegraphics[width=8.25 cm]{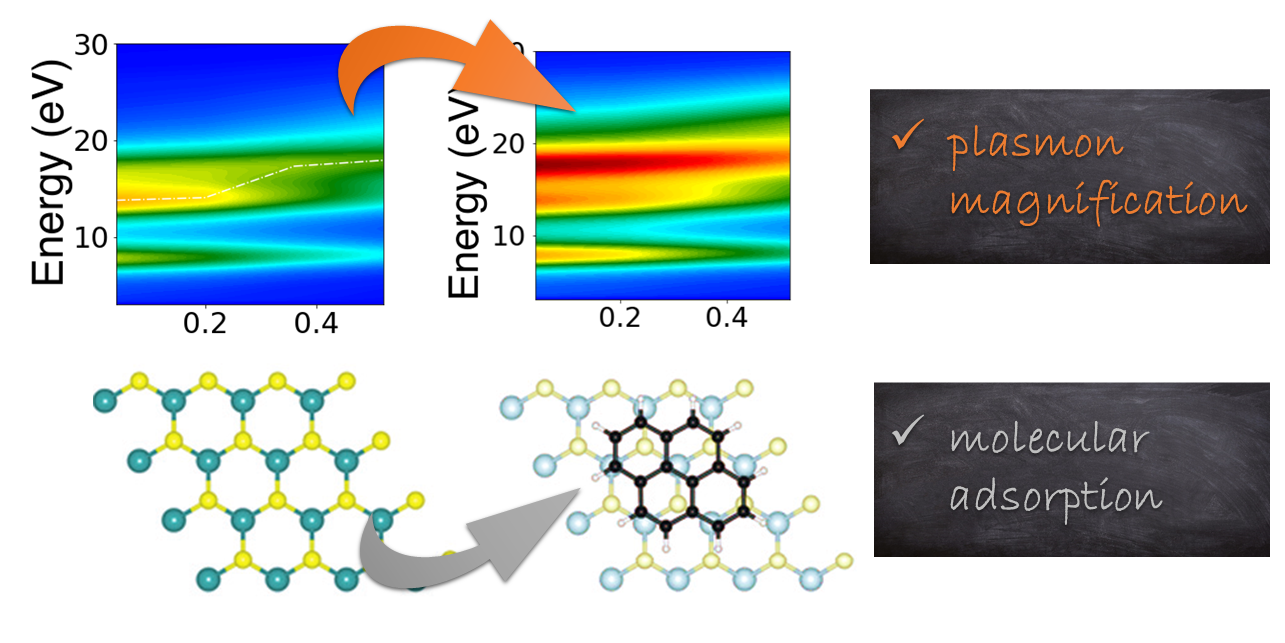}

\end{tocentry}

\begin{abstract}
The adsorption of carbon-conjugated molecules represents an established route to tune the electronic and optical properties of transition metal dichalcogenide (TMDC) monolayers.
Here, we demonstrate from first principles that such a functionalization with prototypical compounds pyrene and tetracene can also enhance the magnitude of selected plasmon resonances in a MoS$_2$ single sheet, without significantly altering their energy and dispersion. Our proof-of-principle results indicate that such a magnification can be achieved by proper alignment of the molecules with respect to the direction of the transferred momentum. The distinct signatures in the loss function of the interface compared to those of its constituents suggest not only the presence of non-negligible interactions between them but also the possibility of using electron energy loss spectroscopy to detect the presence and the orientation of molecular adsorbates on TMDCs.

\end{abstract}

\newpage
\section{Introduction}

Transition metal dichalcogenides (TMDCs) are layered materials held together by van der Waals (vdW) forces~\cite{mars83irpc}. These systems are highly interesting due to their tunable thickness upon exfoliation~\cite{otta+17-2DM,huan+20natcom}, which drastically affects their electronic and optical properties~\cite{mak+10prl,sple+10NL,li+14prb}. 
From a fundamental perspective, TMDCs are characterized by distinct excitonic and plasmonic features~\cite{thyg17-2DM,wang+18rmp}. The latter, in particular, are similar to those of other two-dimensional materials like graphene~\cite{eber+08prb,lu+09prb,kiny+12EL}.
These systems present two primary plasmon resonances, identified as $\pi$ and $\pi+\sigma$ due to the character of the electronic states ruling the corresponding electronic transitions, namely $\pi \rightarrow \pi^*$ and $\sigma \rightarrow \sigma^*$~\cite{mari+04PRB}: $\pi$-resonances are typically found in the near ultraviolet region, between 5 and 10~eV, while $\sigma$-resonances are energetically higher, above 10~eV.
Recent studies based on electron energy loss spectroscopy (EELS) have contributed to shed light on the details of these excitations in TMDCs.
Among the most relevant ones, Moynihan \textit{et. al.} studied the plasmon modes on a pristine single layer of MoS$_2$ \cite{moyn+20JM}, Nerl \textit{et. al.} explored both excitons and plasmons on a few layers of MoS$_2$ by means of EELS~\cite{nerl+17SN}, and Yue and coworkers focused on the $q$-dispersion of plasmons resonances in this material~\cite{yue+17prb}. 

In the last few years, molecular functionalization has emerged as a viable way to modulate the electronic and optical properties of TMDCs~\cite{gobb+18am,huan+18csr,dauk+19apx,ji-choi22ns}. 
The deposition of organic layers on top of these materials gives rise to hybrid interfaces with tunable level alignment~\cite{amst+19nano,park+21as,guo+22nr}, work function~\cite{wang+19aelm,chen+21jpcl}, and spectral response~\cite{kafl+19jacs,chen+20nano} depending on the choice of the constituents.
While organic dopants lead to strong electronic interactions with the substrate~\cite{zhen+16nano,wang+18afm}, thus altering dramatically their intrinsic features already in the ground state, non-polar carbon-conjugated molecules influence the characteristics of the underlying TMDCs with respect to the interaction with electromagnetic fields~\cite{choi+16nano,liu+17nl,xie+19jpca}.
First-principles studies on this class of hybrid materials have contributed to a better understanding of their fundamental properties~\cite{jing+14jmca,cai+16cm,shen-tao17ami,habi+20ats,wang-paul20pccp,krum-cocc21es,mela+22pccp,jaco+22acsanm,soto+22pccp}.
For example, the possibility to computationally explore the effects of a large variety of molecules adsorbed on TMDCs has provided the community not only with a catalog of electronic-structure data but also with a rationale for  predictions~\cite{cai+16cm,mela+22pccp,habi+20ats,zhou+21aplm,slas+23acsaem,soto+22pccp}.
Likewise, the study of the optical absorption properties of such hybrid interfaces, including excitons~\cite{zhan+18am,ulma-quek21nl,oliv+22prm,mark+22nano,decl+23jpcc,thom+23ns}, has contributed to a deeper understanding of these materials and opened new avenues to tune the exceptional optical properties of TMDCs.
On the other hand, the electron energy loss of such systems is still largely unexplored. 
The main reason lies in the technical issues of corresponding measurements which would likely destroy the sample before delivering meaningful data. 
Until a viable solution to this problem is found, first-principles calculations can provide useful references for future experiments and, most importantly, new insight into the fundamental properties of TMDC/molecule interfaces.

In this work, we present a first-principles study of EELS spectra of two prototypical hybrid materials formed by monolayer MoS$_2$ decorated with physisorbed pyrene and tetracene molecules. 
The goal of this study is to understand, in a proof-of-principle fashion, whether and how the loss-function resonances of MoS$_2$ are modified by the presence of physisorbed conjugated molecules. 
As a matter of fact, the chosen carbon-based compounds do not enable the formation of stable interfaces on TMDCs in contrast to larger related molecules already employed in experimental works
\cite{jari+16nl,zhen+16nano,kim+16sr,bett+17nl,yu+17acsel,guo+22nr,tand+23pssa}.
By exploring different configurations in which the moieties are systematically rotated with respect to the direction of the transferred momentum $q$, we analyze the energy, intensity, and dispersion of the plasmon resonances arising in these hybrid interfaces.
We find significant magnification induced by the molecular adsorbates of selected maxima in the loss function of MoS$_2$, which is accompanied by minor variations in terms of energy and $q$-dependent dispersion.
By comparing the results obtained for the interfaces with those of their individual constituents, we discuss the conditions under which the above-mentioned effects occur.
These results provide insight into the interactions between TMDCs and their adsorbates complementary to the existing knowledge mainly based on electronic and optical spectroscopy.
Moreover, the identified enhancement of the plasmon resonances of the TDMC suggests using the loss function as a meaningful observable to detect the presence of physisorbed molecules thereon.

%%%%%%%%%%%%%%%%%%%%%%%%%%%%%%%
\section{Systems and Methods}

\subsection{Pyrene and Tetracene Physisorbed on Monolayer MoS$_2$}

In this study, we consider two prototypical hybrid inorganic/organic interfaces formed by monolayer MoS$_2$ decorated with either pyrene or tetracene molecules adsorbed on its basal plane. 
The choice of these two compounds is suggested on their similar composition (both include only C and H atoms arranged in four phenyl rings), identical symmetry ($D_{2h}$ point group), but different shapes.
Tetracene is a member of the oligoacene family and is characterized by an anisotropic geometry.
Pyrene, in contrast, is more isotropic and is often used in first-principles studies~\cite{krum-cocc21es,mela+22pccp,jaco+22acsanm,oliv+22prm} as a prototype for larger conjugated compounds such as (functionalized) rylenes that favorably adsorb on TMDCs~\cite{yu+17acsel,guo+22nr,zhen+16nano,tand+23pssa}.
These compounds exhibit a weak coupling with MoS$_2$ as extensively discussed in previous work specifically dedicated to this aspect~\cite{mela+22pccp}.
In particular, the band structure and the density of states of the TMDC are negligibly affected by the presence of the physisorbed molecule and, apart from hybridization effects that are ruled by specific conditions~\cite{krum-cocc21es}, the electronic structure of the interface is essentially a superposition of the features of its constituents~\cite{krum-cocc21es,mela+22pccp,jaco+22acsanm,oliv+22prm,qiao+22npj-2dma}.
For this reason, in optical absorption, specific transitions within the molecular frontier orbitals can be targeted, e.g., by a laser pulse~\cite{jaco+22acsanm}.

\begin{figure}
  \centering
  \includegraphics[width=0.5\linewidth]{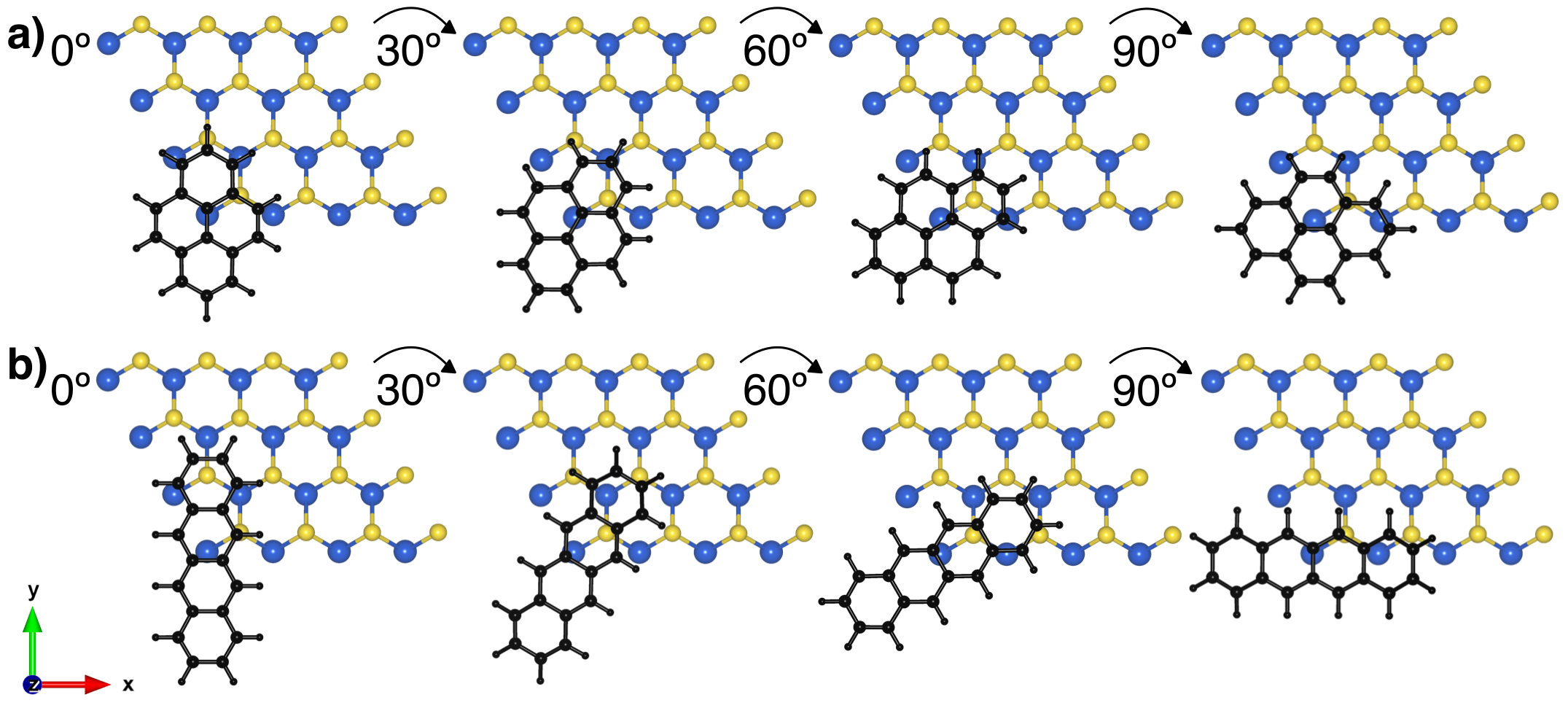}
\caption{Top view of the optimized geometries of a) pyrene and b) tetracene adsorbed on a $4 \times 4$ supercell of MoS$_{2}$ monolayer with different orientations. The $x$-axis, marked by the red arrow, corresponds to the direction of transferred momentum.
}
\label{fig:pyr_tet_geometries}
\end{figure}

In order to explore the impact of the molecular orientation on the loss function of the corresponding interfaces with MoS$_2$, we consider the four different configurations shown in Figure~\ref{fig:pyr_tet_geometries}. 
Rotations with steps of 30$^{\circ}$ are considered, ranging from 0$^{\circ}$, corresponding to the short molecular axis parallel to the $x$-direction marked by a red arrow in Figure~\ref{fig:pyr_tet_geometries}, to the orthogonal configuration 90$^{\circ}$, in which the long molecular axis is aligned with $x$.
Due to the symmetry of both molecules and of the underlying TMDC, rotations of 120$^{\circ}$, 150$^{\circ}$, and 180$^{\circ}$ are equivalent to the geometries at 60$^{\circ}$, 30$^{\circ}$, and 0$^{\circ}$, respectively.

To accommodate pyrene and tetracene on top of a single MoS$_2$ layer, we simulated the TMDC in a $4 \times 4$ hexagonal supercell with lattice parameter $a = 4 \times 3.18$~\AA{} = 12.72~\AA{}. A vacuum layer of 20~\AA{} is included in the out-of-plane direction $z$ to decouple MoS$_2$ from its replicas. 
In the relaxed geometries, the vertical distance between pyrene (tetracene) and the underlying TMDC is 3.3~\AA{} (3.4~\AA{}), in agreement with previous studies on the same or on similar systems~\cite{krum-cocc21es,mela+22pccp,shen-tao17ami,oliv+22prm}.
In the resulting configurations, both molecules form a dense monolayer structure on top of the semiconductor with the spacing between the molecules and their in-plane neighbors depending on their orientation.
In the case of pyrene, the separation from its nearest replica along $x$ and $y$ ranges from 5.9~\AA{} and 5.2~\AA{} in the 0$^{\circ}$ configuration, respectively, to 3.5~\AA{} and 5.7~\AA{} at 90$^{\circ}$.
At intermediate angles of 30$^{\circ}$ and 60$^{\circ}$, the intermolecular separations along $x$ ($y$) are 5.7~\AA{} (3.5~\AA{}) and 4.7~\AA{} (4.7~\AA{}), respectively.
For tetracene, differences among the various orientations are more pronounced due to the anisotropy of this compound.
When the molecule is aligned with its short axis parallel to $x$ (0$^{\circ}$), the distance from its closest replicas in the $x$ and $y$ directions are 7.7~\AA{} and 3.0~\AA{}, respectively.
On the other hand, in the perpendicular configuration (90$^{\circ}$), the relaxed molecules do not lie perfectly flat on the substrate but their opposite ends slightly overlap with each other due to their very short separation (2.1~\AA{}) along their long axes; along the short axes, instead, the distance between nearest replicas is 6.2~\AA{}.
We checked that increasing the spacing among tetracene molecules on MoS$_2$ such that they lay flat on the substrate does not change the results presented below.
For an angle of 30$^{\circ}$, the separation along $x$ ($y$) is equal to 7.2~\AA{} (2.2~\AA{}).
At 60$^{\circ}$, the distance from the neighboring replica is 3.0~\AA{} in both $x$ and $y$ directions.

It is worth stressing that the systems modeled with the procedure described above are not meant to represent realistic interfaces.
First of all, to the best of our knowledge, there is no evidence that small molecules such as pyrene and tetracene can be adsorbed on TMDCs in a stable manner: only larger acenes like pentacene and extended rylenes such as terrylene have been demonstrated to form stable interfaces with TMDCs~\cite{tand+23pssa,kach+21cs,mark+22nano}.
Second, in realistic samples, intrinsic effects such as the density of the molecules on the substrate, the morphology of their clusters, and the thickness of their layers, as well as extrinsic factors such as temperature, pressure, and surface roughness play a decisive role in determining the characteristics of the interface.
All these aspects should be carefully addressed for a reliable comparison with experiments but they go beyond the scope of the present work.

%%%%%%%%%%%%%%%%%%%%%%%%%%%%%%%%
\subsection{Computational Methods}
The results presented in this work are obtained in the framework of density functional theory~\cite{hohe-kohn64pr} and its time-dependent extension~\cite{rung-gros84prl}. 
Ground-state properties are calculated by solving the Kohn-Sham (KS) equations~\cite{kohn-sham65pr}, while EELS spectra are computed in linear response using the Lanczos algorithm~\cite{malc+11cpc}.
This approach is widely established in the community and has been applied to various classes of materials including elemental crystals~\cite{timr+17prb}, binary compounds~\cite{cuda-wirt21prb}, and interfaces~\cite{jia+21nanotech}. 
The key output quantity of these calculations is the loss function, defined as $L(\mathbf{q},\omega) = - \text{Im}\,\varepsilon^{-1}(\mathbf{q},\omega)$, where $\varepsilon$ is the macroscopic dielectric function of the material evaluated from its polarizability $\chi$.
The latter quantity represents the charge-density response function calculated from the solutions of the KS equations including the adopted approximation for the exchange-correlation functional.
For further details on this approach, we redirect interested readers to a specialized review~\cite{onid+02rmp}.
The dispersion of the loss function is simulated considering non-zero values of transferred momentum $\textbf{q}$ in the in-plane direction $\Gamma$-M, where M corresponds to the mid-point at the edge of the two-dimensional hexagonal Brillouin zone according to the usual convention~\cite{cast+09rmp}. 
Assuming the direction of momentum transfer along the $x$-axis, only the corresponding component of the vector $\textbf{q}$ ($q_x$) is varied.
Four values of $q$ ranging from 0.04~$(2\pi/a)$, approaching the optical limit at $q \rightarrow 0$, to 0.52~$(2\pi/a)$ are considered. 
This range of momentum transfer was chosen based on preliminary results on an isolated MoS$_2$ monolayer, indicating that above $q = 0.52 \, (2\pi/a)$ the features of the loss function in this material are strongly smeared and, hence, hardly recognizable.

All calculations are performed with Quantum ESPRESSO~\cite{gian+17jpcm} adopting the SG15 Optimized Norm-Conserving Vanderbilt pseudopotentials~\cite{hama13prb} and a plane-wave basis set with kinetic energy and charge-density cutoff of 60 Ry and 300 Ry, respectively.
The Perdew-Burke-Ernzerhof~\cite{perd+96prl} functional is employed to approximate the exchange-correlation potential supplemented by the Tkatchenko-Scheffler scheme~\cite{tkat-sche09prl} to account for pairwise vdW interactions between molecules and substrate. 
The Brillouin zone of the hybrid interface and of the isolated TMDC simulated in the same supercell shown in Figure~\ref{fig:pyr_tet_geometries} is sampled by a $2 \times 2 \times 1$ $\mathbf{k}$-mesh. 
The loss function of the freestanding molecules is calculated for consistency in the same $4 \times 4$ supercell and with the moieties in the same arrangement formed in each heterostructure according to the rotation angle with respect to $x$.
In these calculations, a denser $6 \times 6 \times 1$ \textbf{k}-grid is adopted to better resolve the spectral features of the molecules.
The considered hybrid interfaces are optimized with the quasi-Newtonian Broyden-Fletcher-Goldfarb-Shanno algorithm~\cite{broy70imajam,flet70cj,gold70mc,shan70mc} with force and energy thresholds of $10^{-3}$ Ry/bohr and $10^{-4}$ Ry, respectively.  
The convergence of the EELS spectra is obtained with $500$ Lanczos coefficients in the underlying algorithm, which are effectively extrapolated up to $40000$ in post-processing.
%%%%%%%%%%%%%%%%%%%%

\section{Results and Discussion}

\subsection{Loss Function of Pyrene@MoS$_2$}
\begin{figure}[h!]
  \centering
  \includegraphics[width=0.5\linewidth]{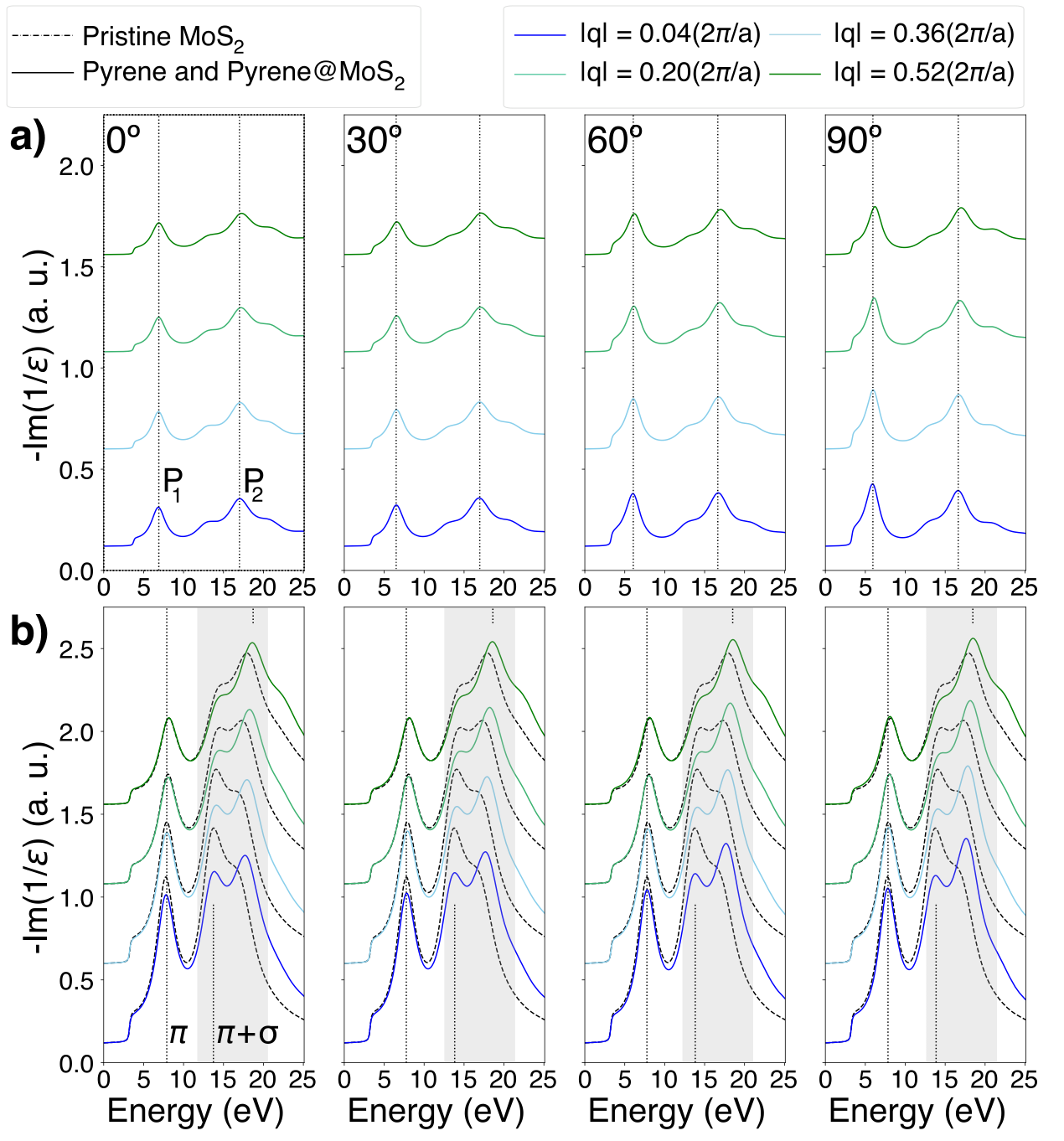}
\caption{Electron energy loss spectra at different values of transferred momentum $q$ of a) freestanding pyrene, b) pristine MoS$_2$ (dashed curves), and pyrene@MoS$_{2}$ (solid curves). The gray area in panel b) highlights the energy range in which the most significant changes to the loss function of MoS$_2$ occur due to molecular physisorption. The spectra computed at increasing values of $q$ are offset for better visibility. %A broadening of 100 meV is applied to all spectra.
}
\label{fig:pyr_MoS2_LF}
\end{figure}

%\label{sec:pyr}
We start our analysis from the loss functions of the hybrid interface pyrene@MoS$_2$, see Figure~\ref{fig:pyr_MoS2_LF}.
For comparison, we examine in parallel the results obtained for the freestanding molecular layer in the same arrangement as in the heterostructures (Figure~\ref{fig:pyr_MoS2_LF}a) and the pristine TMDC without adsorbates (Figure~\ref{fig:pyr_MoS2_LF}b, dashed curves).
This strategy is most appropriate for the analysis of the following results, given the broad energy range spanned by the calculated EELS spectra and the non-negligible influence of momentum transfer, which cannot be appropriately captured by standard electronic-structure calculations.

The EELS spectrum of pyrene is characterized by two main peaks at about 7~eV and 17~eV, labeled P$_1$ and P$_2$, respectively. 
Energy and relative intensity of these maxima are in agreement with experimental data collected for pyrene in vapor phase~\cite{koch-otto69oc,veng-hinz76jcp}, which ascribe the former to the manifold of $\pi$-$\pi^*$ transitions~\cite{bito+00cpl,mall+04aa} and the latter to $\sigma$-$\sigma^*$ transitions~\cite{mall+04aa}.
The characteristics of the peaks do not vary significantly either with increasing values of $q$ or upon rotation of the molecules.
However, examining the results in more detail, one notices that the lowest-energy maximum becomes sharper and slightly red-shifted as the short molecular axis forms an angle of 90$^\circ$ with respect to the $x$ direction (see Figure~\ref{fig:pyr_MoS2_LF}a).
This is due to the fact that depending on the orientation of the molecule, excitations with orthogonal polarization contribute to the peaks in the EELS.
The results obtained at 0$^\circ$ and 90$^\circ$  can be compared with polarization-resolved EELS measurements performed on gaseous pyrene~\cite{veng72pssb}. 
Like our findings, the experimental data show a slight increase in the relative intensity of the first resonance when the molecule has its long axis aligned with $x$.

In the EELS of the pyrene@MoS$_2$ interface (solid curves in Figure~\ref{fig:pyr_MoS2_LF}b), we identify the $\pi$-resonance of MoS$_2$ around 8~eV~\cite{nerl+17SN,moyn+20JM} (dashed curves in Figure~\ref{fig:pyr_MoS2_LF}b), which is clearly visible also for the interface.
At 14~eV, another sharp peak appears in the loss function of the TMDC monolayer followed by a shoulder at approximately 16~eV: these features belong to the $\pi+\sigma$ band~\cite{moyn+20JM}.
Upon increasing momentum transfer, the main variation in the spectrum of MoS$_2$ is the intensity redistribution between these two maxima.
For $q=0.36~(2\pi/a)$, the two features have almost equivalent strength while for larger values of $q$ the higher-energy one becomes more intense (see dashed curves in Figure~\ref{fig:pyr_MoS2_LF}b). 
The effects of pyrene physisorption manifest themselves mostly in the region 12-20~eV, which is highlighted in gray in Figure~\ref{fig:pyr_MoS2_LF}b.
The $\pi$-resonance is negligibly influenced by the presence of the molecule regardless of its orientation.
On the other hand, the $\pi+\sigma$ band is largely affected by adsorbed pyrene.
At $q \rightarrow 0$, the relative spectral weight of the two maxima is reversed compared to the result obtained for the isolated TMDC (compare solid and dashed curves in Figure~\ref{fig:pyr_MoS2_LF}b).
A blue shift of $\sim$1-2~eV is additionally noticed for both maxima, as expected in light of their plasmonic nature~\cite{nerl+17SN}.
The second peak in the $\pi+\sigma$ band remains more intense than the first one even at increasing transferred momentum, such that the loss function computed for the hybrid interface at $q=0.52~(2\pi/a)$ resembles quite closely the one obtained for MoS$_2$ with the same momentum transfer.
The only visible difference is again the blue shift of the second resonance anticipating the dispersion discussed in detail below.
The impact of pyrene orientation on the EELS is very moderate, consisting essentially of a slight, relative increase in the strength of the peak at 19~eV as the molecule approaches an angle of 90$^\circ$ with respect to the $x$ axis (see Figure~\ref{fig:pyr_MoS2_LF}b).

%%%%%%%%%%%%%%%%%%%%%%%%%%%%%%%%%%%%%
\subsection{Loss Function of Tetracene@MoS$_2$}
\begin{figure}[h!]
  \centering
  \includegraphics[width=0.5\linewidth]{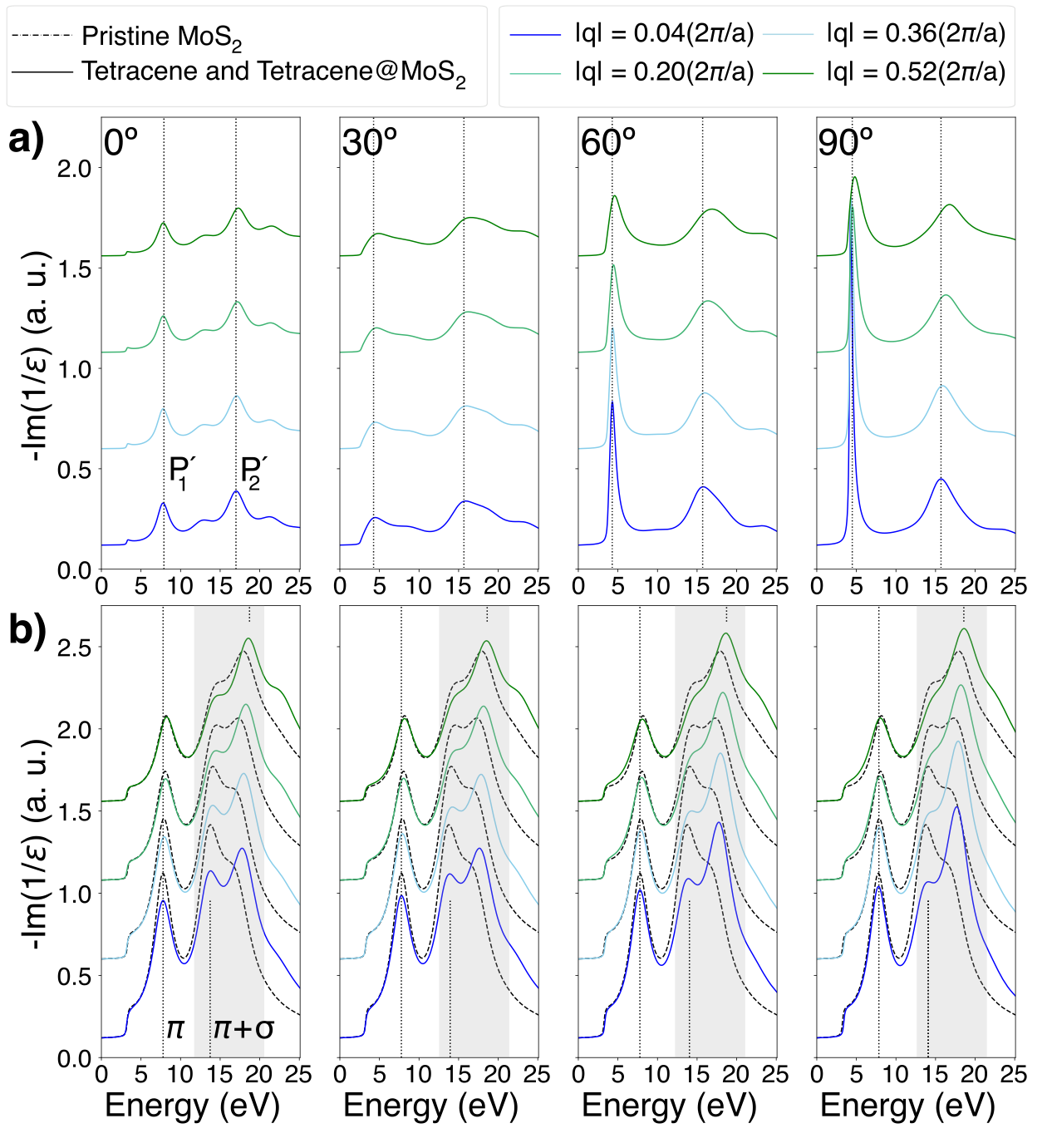}
\caption{Electron energy loss spectra at different values of transferred momentum $q$ of a) freestanding tetracene, b) isolated MoS$_2$ (dashed curves), and tetracene@MoS$_{2}$ (solid curves). The gray area in panel b) highlights the energy range in which the most significant changes to the loss function of MoS$_2$ occur due to the physisorbed molecules. The spectra computed at increasing values of $q$ are offset for better visibility. 
}
\label{fig:tet_MoS2_LF}
\end{figure}

We now turn to the analysis of the loss function calculated for the tetracene@MoS$_2$ interface, see Figure~\ref{fig:tet_MoS2_LF}.
Following the same pattern adopted above for pyrene@MoS$_2$, we inspect the spectra of the isolated components as a basis to understand the behavior obtained for the hybrid system.
The $q \rightarrow 0$ EELS spectrum of the freestanding tetracene layer at 0$^\circ$ (short molecular axis parallel to $x$, see Figure~\ref{fig:pyr_tet_geometries}b) is characterized by two maxima at about 8~eV (P$_1^{'}$) and 17~eV (P$_2^{'}$), see Figure~\ref{fig:tet_MoS2_LF}a.
P$_1^{'}$ corresponds to an optical excitation polarized along the short axis of tetracene that is visible also in its absorption cross section~\cite{mall+04aa}.
Notice that this is not the lowest-energy excitation of the tetracene molecule, which is also polarized along the short axis but appears at lower energy, around 2~eV~\cite{mall+07cp,mall+11cp}.
The overall similarity with the loss functions of pyrene (Figure~\ref{fig:pyr_MoS2_LF}a) is due to the common features exhibited by all carbon-based networks~\cite{kell+92apj,yin-zhan12jap} including graphene~\cite{mari+04PRB}.
However, in contrast to pyrene and as a consequence of its more pronounced structural anisotropy, the EELS of tetracene is very sensitive to the orientation of the molecule with respect to the direction of transferred momentum.
When the molecule is oriented with its long axis parallel to $x$, the loss function is dominated by a very sharp resonance at about 5~eV and by a broader but still pronounced maximum at $\sim$16~eV.
The former corresponds to the second bright excitation in the spectrum of tetracene polarized along its long molecular axis~\cite{mall+11cp}.
%The substantially larger intensity of this peak compared to the first one at 0$^\circ$ can be understood considering the dependence of the transition dipole moment on the distance between the domains with opposite charge~\cite{cocc+12jpcl,cocc+14jpca}.
As mentioned, this is not the lowest-energy transition in this molecule, which is instead polarized along the short axis, as in all oligoacenes~\cite{humm-ambr05prb,cocc+22jpm}.
At intermediate orientations of tetracene, the $q \rightarrow 0$ EELS contains a combination of the signatures dominating the signals at 0$^\circ$ and 90$^\circ$.
At 60$^\circ$, the sharp resonance close to 5~eV is still clearly evident, although less intense than at 90$^\circ$, while the second maximum is broader and red-shifted.
Finally, at 30$^\circ$, the two peaks are significantly smeared out but the maxima are energetically very close to P$_1^{'}$ and P$_2^{'}$ at 0$^\circ$.
Notably, the sharp resonance at ~$\sim$5~eV dominating the EELS of tetracene oriented with its long axis parallel to $x$ is detectable experimentally~\cite{koch-otto69oc} like the weaker peak above 15~eV. 
The main features in the loss function of tetracene in all considered orientations are generally preserved in energy and intensity independently of $q$ (see Figure~\ref{fig:tet_MoS2_LF}a).
The only visible exception is at 90$^\circ$ and to a lesser extent at 60$^\circ$, where the strength of the first resonance decreases with increasing momentum transfer.

Examining now the EELS spectrum of the tetracene@MoS$_2$ interface (Figure~\ref{fig:tet_MoS2_LF}b), we notice an overall similarity with the loss function of pyrene@MoS$_2$.
The orientation of the molecule and the magnitude of $q$ have little influence on the main spectral features.
The low-energy part of the EELS is dominated by the $\pi$ resonance of MoS$_2$.
Even when the long axis of the molecule is aligned along $x$ (90$^\circ$), the relative intensity of the first peak increases only slightly with respect to the $\pi$ resonance in the isolated TMDC.
Only a very careful inspection of Figure~\ref{fig:tet_MoS2_LF}b and Figure~\ref{fig:pyr_MoS2_LF}b reveals that this enhancement is larger in the presence of tetracene than of pyrene. 
On the other hand, the variations induced by the adsorbed molecule on the $\pi + \sigma$ resonance are even more dramatic than in the spectrum of pyrene@MoS$_2$, especially for 60$^\circ$ and 90$^\circ$.
In these cases, the relative intensity of the peak at $\sim$19~eV is considerably magnified compared to the corresponding feature in the EELS of isolated MoS$_2$. 
Notably, the higher intensity of this maximum in the loss function of the hybrid interface is clearly visible also at large values of $q$, such that the resemblance with the spectrum of the pristine monolayer is much less pronounced than in EELS of tetracene@MoS$_2$ in the region 15-20~eV highlighted in gray in Figure~\ref{fig:tet_MoS2_LF}b.

%%%%%%%%%%%%%%%%%%%%%%%%%%
\subsection{Energy Dispersion of the Loss Function}
\begin{figure}
  \centering
  \includegraphics[width=0.85\linewidth]{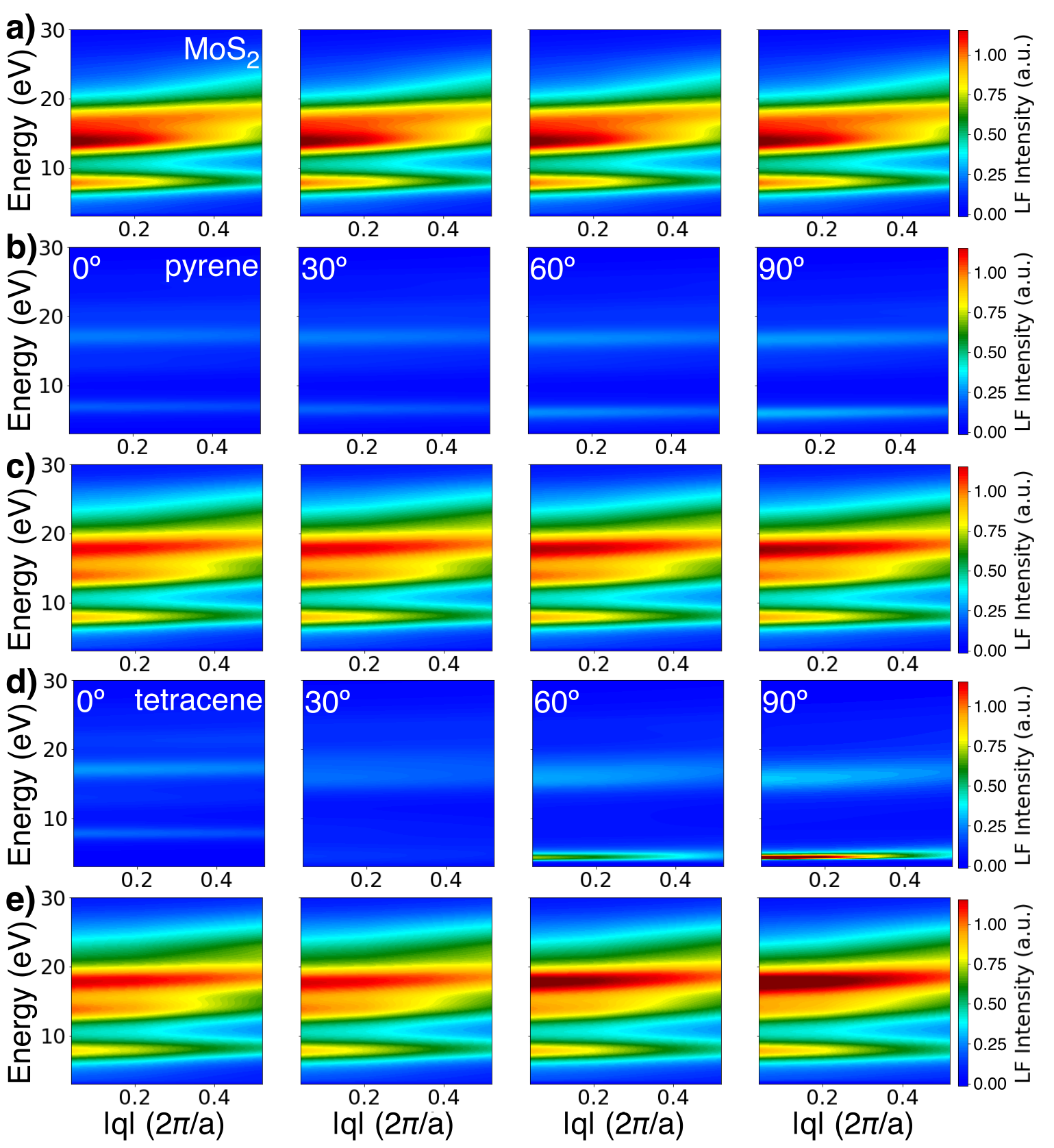}
\caption{Energy dispersion of the loss function (LF) of a) the isolated MoS$_2$ monolayer, b) freestanding pyrene, c) pyrene@MoS$_2$, d) freestanding tetracene, and e) tetracene@MoS$_2$. The molecular orientations indicated in panels b) and d) hold also for panels c) and e). }
\label{fig:pyr_tet_ColorPlot}
\end{figure}

We conclude our analysis by discussing the energy dispersion of the loss function, which can be carried out most conveniently by plotting the EELS as a function of energy and momentum simultaneously.
The graphs reported in Figure~\ref{fig:pyr_tet_ColorPlot} contain the same information shown in Figures~\ref{fig:pyr_MoS2_LF} and \ref{fig:tet_MoS2_LF} but they offer a clearer visualization of the EELS dispersion in the hybrid heterostructures compared to their freestanding constituents.
The $\pi$ resonance dominating the loss function of the isolated MoS$_2$ is clearly visible in Figure~\ref{fig:pyr_tet_ColorPlot}a, where the decreasing intensity of this feature at increasing momentum transfer is apparent.
In contrast, the $\pi+\sigma$ band, which is extended over an energy range of about 8~eV, exhibits a positive dispersion, especially at its boundaries at 12 and 20~eV in the spectrum at $q \rightarrow 0$ (cyan curves in Figure~\ref{fig:pyr_tet_ColorPlot}a).
Remarkably, other stronger maxima that are visible at 12 and 20~eV in the optical limit, are dispersive, too (light green in Figure~\ref{fig:pyr_tet_ColorPlot}a).
However, their increase in energy as a function of $q$ is not as steep as for the above-mentioned lateral peaks. 
A similar trend is exhibited also by the intense feature centered at 17~eV.
On the other hand, the strongest maximum at 14~eV is almost non-dispersive: at increasing $q$, only its intensity decreases, as discussed above, while its energy remains almost constant.

Finally, we inspect the EELS dispersion, which can be best appreciated in the color plots shown in Figure~\ref{fig:pyr_tet_ColorPlot}.
We start from the freestanding molecular layers, see Figure~\ref{fig:pyr_tet_ColorPlot}b and d for pyrene and tetracene, respectively.
In the case of pyrene, the two maxima, P$_1$ and P$_2$, appear as dispersionless features, independent of the orientation of the moieties. 
The intensity of both spectral features is only moderately changing with respect to the rotation angle: the strongest signal is obtained when the long molecular axis is parallel to $x$ (90$^{\circ}$), due to the larger oscillator strength of the corresponding electronic excitations along that direction~\cite{bito+00cpl,herp+21jpca}.
Accordingly, the maximum at 90$^{\circ}$ is slightly below in energy compared to the one at 0$^{\circ}$, as discussed with reference to Figure~\ref{fig:pyr_MoS2_LF}a.
The situation is rather different in the case of tetracene, see Figure~\ref{fig:pyr_tet_ColorPlot}d, where the high anisotropy of this molecule leads to significantly different signals depending on its orientation with respect to the transferred momentum. 
For tetracene aligned with the short axis parallel to $x$ (0$^{\circ}$), two dispersionless features are seen at about 8~eV and 17~eV. 
A higher-energy maximum above 20~eV, which is visible as a shoulder in Figure~\ref{fig:tet_MoS2_LF}b, can be noticed, too.
At 30$^{\circ}$, only the resonance around 17~eV is visible, while the lower-energy one is too weak to be detected with the adopted color scale. 
At 60$^{\circ}$ and 90$^{\circ}$, the sharp maximum at about 4.5~eV gives rise to an intense signal in the color plot, which is strongest when the long molecular axis is aligned along $x$ (see Figure~\ref{fig:pyr_tet_ColorPlot}d, middle panels).
Due to the localized character of this electronic excitation, the corresponding maximum in the loss function is dispersionless.
Conversely, the feature at 17~eV exhibits a hint of a positive dispersion, which is most evident at 90$^{\circ}$ but visible also at 60$^{\circ}$.
We speculate that this behavior is due to the one-dimensional, carbon-conjugated lattice effectively generated by the molecule and its close-by replicas, where the electronic charge can delocalize itself.
The displayed feature is compatible with the plasmonic behavior of a bilayer structure~\cite{poli-chia14ns,moyn+20JM}, suggesting that the molecular layer effectively increases the thickness of the system compared to MoS$_2$ alone.

With the understanding gained of the $q$-dependent loss function of the individual building blocks, we finally analyze the results obtained for the hybrid interfaces.
The EELS of pyrene@MoS$_2$ as a function of $q$ is dominated by two main features (Figure~\ref{fig:pyr_tet_ColorPlot}c).
The one centered at approximately 8~eV corresponds to the $\pi$ resonance, which is dispersionless and exhibits a decreasing intensity at increasing values of $q$ as in the isolated TMDC.
However, the intensity of the $\pi$ band at $q \rightarrow 0$ is lower in the hybrid system compared to MoS$_2$ alone (compare Figure~\ref{fig:pyr_tet_ColorPlot}a and c).
We speculate that the presence of physisorbed pyrene, which features a maximum in the EELS in the close vicinity of the $\pi$ resonance of the TMDC (Figure~\ref{fig:pyr_tet_ColorPlot}b) partially suppresses the intensity of this resonance.
The fact that the strength of the $\pi$ band in the interface with the molecules at 90$^{\circ}$ is slightly larger toward low values of $q$ compared to the system at 0$^{\circ}$ reinforces this hypothesis, given the corresponding increase in intensity of P$_1$ seen in the corresponding EELS plots (in Figure~\ref{fig:pyr_tet_ColorPlot}b).

The $\pi+\sigma$ band is even more dramatically affected by the presence of the organic adsorbates. 
The influence of the molecules manifests itself in two ways.
On the one hand, the relative intensity of the two main maxima at 14 and 19~eV is reverted and this effect is more pronounced when the molecules are aligned with their long axes parallel to the direction of the transferred momentum (90$^{\circ}$).
Notably, both features remain essentially dispersionless as in the spectrum of MoS$_2$ alone (see Figure~\ref{fig:pyr_tet_ColorPlot}a).
On the other hand, there is a slight blue shift accompanied by a broadening of the entire $\pi+\sigma$ resonance, especially toward higher energies. 
This behavior can be interpreted again as a consequence of the effective bilayer nature of the system~\cite{moyn+20JM, poli-chia14ns} owing to the presence of the molecular monolayer on top of MoS$_2$. 

The dispersion relation of the loss function of the tetracene@MoS$_2$ interface (Figure~\ref{fig:pyr_tet_ColorPlot}e) exhibits similar features as those of pyrene@MoS$_2$: The $\pi$ resonance is dispersionless and undergoes a significant decrease in intensity at increasing values of $q$, while the $\pi + \sigma$ band, dominated by two intense maxima at 14 and 19~eV, exhibit a clear positive dispersion, especially at its extrema.
This being said some noteworthy differences emerge due to the more pronounced molecular anisotropy of tetracene.
The increase in the intensity of the second resonance in the $\pi + \sigma$ band (at $\sim$19~eV) is significantly more pronounced than in the hybrid interface between MoS$_2$ and pyrene. 
We can relate this behavior to the substantial magnitude gain of P$_2^{'}$ when the molecule is at 90$^{\circ}$ rather than at 0$^{\circ}$ (Figure~\ref{fig:pyr_tet_ColorPlot}d).
On the other hand, the very strong maximum exhibited by the EELS of tetracene at 90$^{\circ}$ at about 4~eV does not participate in the EELS of the hybrid interface: there is no signal at that energy in Figure~\ref{fig:pyr_tet_ColorPlot}e.

%%%%%%%%%%%%%%%%%%%%%%%%%%%%%%%
\section{Conclusions}
In summary, we have investigated the $q$-dependent loss function of two prototypical hybrid interfaces formed by the carbon-conjugated molecules pyrene and tetracene physisorbed on single-layer MoS$_2$. 
To assess the role of the molecular orientation with respect to the momentum transfer, we have considered four different configurations with the organic components rotated in steps of 30$^{\circ}$.
The loss function of the hybrid systems is dominated by the $\pi$ and $\pi + \sigma$ resonances characterizing the response of the isolated TMDC monolayer.
While the $\pi$ band, around 7-8~eV in both systems, is almost unaffected by the presence of the molecule, the intensity and the energy of the $\pi + \sigma$ resonance are largely influenced by the adsorbates. 
In particular, we see a magnification of the sub-peaks in the latter in comparison to the spectrum of the pristine TMDC. 
This effect is more evident at low values of transferred momentum, although they are visible also at high $q$.
The larger anisotropy of tetracene compared to pyrene further enhances this behavior when the molecule is aligned with its long axis parallel to the incoming momentum. 
The dispersion of the loss function provides additional information about the nature of identified features. 
While the $\pi$ resonance is dispersionless and its strength decays with increasing $q$, in agreement with the result obtained for the isolated MoS$_2$ monolayer, the $\pi + \sigma$ band is characterized by a more complex behavior.
The weaker maxima at its extrema exhibit a positive dispersion resembling the trend of multilayer structures~\cite{poli-chia14ns}.
Moreover, the increasing intensity of these features in the presence of the molecule is a signature of the coupling between the organic and inorganic constituents of the interface.

In conclusion, our results reveal in a proof-of-principle fashion that the physisorption of conjugated molecules enhances selected resonances in the $q$-dependent loss function of TMDCs, due to the energetic proximity of corresponding maxima in the isolated compounds. 
This magnification is a signal of the electronic interactions between the constituents which can be exploited to detect the presence of organic adsorbates on the inorganic substrate.
The dependence on the orientation of the molecules with respect to the direction of the transferred momentum can be further used to identify the structural arrangement of the adsorbates.
Future EELS experiments on such systems will be very helpful to confirm our results or to stimulate new theoretical investigations.
In this case, a detailed analysis of the morphology of the sample and of the effects of temperature will be necessary to ensure a reliable comparison with the experimental data.

%%%%%%%%%%%%%%%%%%%%%%%%%%%%%%%%%%%%
\begin{acknowledgement}
The authors are grateful to Jannis Krumland and M. Sufyan Ramzan for their insightful comments on the unpublished manuscript. 
This work was funded by the German Research Foundation, Project No. 182087777 - CRC 951, by the German Federal Ministry of Education and Research (Professorinnenprogramm III) as well as from the State of Lower Saxony (Professorinnen f\"ur Niedersachsen, DyNano, and SMART). Computational resources were provided by the North-German Supercomputing Alliance (HLRN), project bep00104, and by the high-performance computing cluster CARL at the University of Oldenburg, funded by the German Research Foundation (Project No. INST 184/157-1 FUGG) and by the Ministry of Science and Culture of Lower Saxony.
\end{acknowledgement}
%%%%%%%%%%%%%%%%%%%%%%%%%%%

%%%%%%%%%%%%%%%%%%%%%%%%%%%%%%%%%%%%
%\bibliography{biblio}

\providecommand{\latin}[1]{#1}
\makeatletter
\providecommand{\doi}
  {\begingroup\let\do\@makeother\dospecials
  \catcode`\{=1 \catcode`\}=2 \doi@aux}
\providecommand{\doi@aux}[1]{\endgroup\texttt{#1}}
\makeatother
\providecommand*\mcitethebibliography{\thebibliography}
\csname @ifundefined\endcsname{endmcitethebibliography}
  {\let\endmcitethebibliography\endthebibliography}{}

\end{document}